# A Minimized Mutual Information retrieval for simultaneous atmospheric pressure and temperature

*Prabhat K. Koner and James R. Drummond*

## Abstract

The primary focus of the Mars Trace Gas Orbiter (TGO) collaboration between NASA and ESA is the detection of the temporal and spatial variation of the atmospheric trace gases using a solar occultation Fourier transform spectrometer. To retrieve any trace gas mixing ratios from these measurements, the atmospheric pressure and temperature have to be known accurately. Thus, a prototype retrieval model for the determination of pressure and temperature from a broadband high resolution infrared Fourier Transform spectrometer experiment with the Sun as a source on board a spacecraft orbiting the planet Mars is presented. It is found that the pressure and temperature can be uniquely solved from remote sensing spectroscopic measurements using a Regularized Total Least Squares method and selected pairs of micro-windows without any a-priori information of the state space parameters and other constraints.

The selection of the pairs of suitable micro-windows is based on the information content analysis. A comparative information content calculation using Bayes theory and a hyperspace formulation are presented to understand the information available in measurement. A method of minimization of mutual information is used to search the suitable micro-windows for a simultaneous pressure and temperature retrieval.

## Introduction

In order for atmospheric temperature and pressure to be inferred from measurements from a Fourier Transform Spectrometer orbiting Mars, the source of emission must be a relatively abundant gas of known and uniform distribution. Otherwise, the uncertainty in the abundance of the gas will make the determination of temperature and pressure from the measurements ambiguous. Fortunately carbon dioxide is present in the Martian atmosphere in uniform abundance for altitudes below about 200 km, and has emission bands in spectral regions that are convenient for measurement.

There are many papers and books that review the retrieval theories which have been developed over the past few decades [1-5].  The consensus is that retrieving an atmospheric profile of (for example) temperature and pressure from spectroscopic measurements from space is an ill-posed problem for which there is no unique solution because (a) the outgoing radiances arise from relatively deep layers of the atmosphere,

(b) the radiances observed within various spectral channels come from overlapping layers of the atmosphere and are not vertically independent of each other, and (c) the measurements have errors. As a consequence a large number of approaches have been tried. They differ both in the procedure for solving the set of spectrally independent radiative transfer equations (e.g., matrix inversion, numerical iteration) and in the type of ancillary data used to constrain the solution to ensure a meaningful result (e.g., the use of atmospheric covariance statistics, hydrostatic constraint and an a priori estimate of the profile structure etc.). We have already reported in previous papers [6-8] that it is possible to obtain unique solution under some circumstances with proper attention to the mathematics, proper experiment design and reasonable assumptions in the forward model. We show in this paper that pressure and temperature can be uniquely retrieved from the spectroscopic measurement of the Martian atmosphere.

There are two main geometries for these types of measurement: nadir (downward-looking) and limb viewing. Nadir viewing can rely on atmospheric emission, surface emission or reflected solar radiation depending upon the experiment design. Limb viewing uses atmospheric emission or a stellar/lunar source, usually the sun. Nadir techniques have the potential for good horizontal spatial localisation, but worse vertical resolution whereas limb techniques, particularly solar occultation, have the reverse characteristics: potentially high vertical resolution and poor horizontal localisation. The technique discussed in this paper is solar occultation, chosen for its high sensitivity and vertical resolution.

As mentioned above there are many pressure and temperature retrieval algorithms [9-13] available for spectroscopic measurements in the Earth's atmosphere. They frequently use parameter reduction with the help of the hydrostatic equation or some other relationship and a-priori information. The accuracy of the tangent height – the lowest altitude of the limb path - plays an important role when the problem is constrained by the hydrostatic equation. In the case of Mars it will be more difficult to get good pointing knowledge relative to the surface because of the difficulty of knowing the orbit and relevant topography precisely. However the relative nature of the tangent heights in a series of measurements through the satellite orbit is determined mainly by time and the satellite orbital velocity, both of which are very well-known. Thus the relative nature of the tangent heights will be well-known, but the absolute nature less well.

There are few papers on spectroscopic measurements of the Martian atmosphere. Recently a paper [14] reported on the retrieval of pressure and temperature of Mars from the Mars Climate Sounder (MCS) measurement on the Mars Reconnaissance Orbiter using the Chahine method to get an approximate retrieval. MCS is a nadir-viewing instrument. The paper reported that there is no unique solution for the problem and the results are heavily dependent on the initial guess/constrained profile.

An additional consideration is that dust opacity is significant in the Martian atmosphere to an extent not seen upon Earth and the atmospheric temperature profile of Mars can be strongly influenced by the suspended dust.

Finally, the a-priori information is very poor simply due to our ignorance of the Martian atmosphere.

This paper presents a methodology to solve the radiative transfer retrieval problem in a finite parametric space for an orbiting solar occultation spectrometer probing the Martian atmosphere considering pressure and temperature as two independent variables, no a-priori profiles and no hydrostatic constraint.

## *Theoretical Background*

The information content of a datum is a relative measure and has quantitative meaning only in a defined framework. The information content of data can be considered with respect to models if compatible models exist. Radiative transfer modelling for atmospheric problem using line by line (LBL) calculation is mature. The model consists of a set of targeted parameters, which is embedded in a mathematical framework. A model containing "N" parameters can be described by a point in a N-dimensional space. The information gained in a measurement can be quantified by the reduction of the volume of the region in hyperspace [15]. The information, $H_{hs}$, in bits, required to reduce region 1 of volume $R_1$ to region 2 of volume $R_2$ can be written

$$H_{hs} = \log\left(\frac{R_1}{R_2}\right) \tag{1}$$

Since the amount of information is necessarily finite, the regions in hyperspace defined by the model cannot be infinitely large or infinitely small and "truth" has to be found within the region.

The evaluation of the volume of R in a high dimensional space is difficult. If we assume that the points are normally distributed in hyper-ellipsoidal surface [16] then

$$\mathbf{x}^T \mathbf{V}^{-1} \mathbf{x} = c \tag{2}$$

where $\mathbf{x}$ is any displacement vector, $\mathbf{V}$ is the covariance matrix for the parameters, and c is a constant that defines a particular equal-probability surface. The "volume" R of the region bounded by the surface defined by Eqn(2) is [15,17]

$$R = \frac{\sqrt{\pi} \, c^2}{\Gamma(N/2+1)} \sqrt{|\mathbf{V}|} \tag{3}$$

where $\Gamma$ is the gamma function, $|.|$ is the determinant of the matrix , and N is the dimensionality of the space or variables.

Combining Eqn(1) & Eqn(3), $H_{hs}$ can be written as

$$H_{hs} = \frac{1}{2}\log_2\left|\frac{\mathbf{V_1}}{\mathbf{V_2}}\right| \tag{4}$$

This is same formula as that obtained directly from the formulation of Shannon [18] when applied to a distribution of estimators that is multivariate normal. For example, the measure of entropy for the parameter space **x**, i.e.

$$\langle \log[f(x)] \rangle = \int f(x) \log[f(x)] dx \qquad (5)$$

where $\langle . \rangle$ is the expected value and $f(x)$ is the probability density function for the model parameters. The information content of an experiment can then be defined as the entropy based on the prior distribution of x minus the entropy based on the posterior distribution (i.e., that based on applying Bayes' rule to the observed data and the prior distribution). Under multivariate normality, the entropy is a constant +(1/2) [log(det V)], where V is the covariance matrix for the distribution of x. Thus, if $V_1$ and $V_2$ are equivalent to the prior and posterior covariances, respectively, a result identical to Eqn (4) can be obtained.

For a forward model $\mathbf{y} = \mathbf{Kx} + \delta$, the covariance matrix V is usually computed $\mathbf{V} = \delta^2 (\mathbf{K}^T \mathbf{K})^{-1}$. Where $\mathbf{K}$ is forward operator and $\delta$ is the estimate of the variance of the observation errors. $\mathbf{K}$ is conventionally called the Jacobian matrix for the linear and vector valued function/equation. For the assumption that the error in all measurements is the same, $H_{hs}$ can also be defined in terms of K as

$$H_{hs} = \frac{1}{2} \log_2 \frac{\left| \mathbf{K_2}^T \mathbf{K_2} \right|}{\left| \mathbf{K_1}^T \mathbf{K_1} \right|} \qquad (6)$$

The determinant of $\mathbf{K}^T \mathbf{K}$ approaches zero when the condition number of $\mathbf{K}^T \mathbf{K}$ is high, which may produce an inappropriate result. In such a situation, regularization is introduced. Thus, a better assumption of covariance is the inverse of the regularized Jacobian under a least squares formulation, which is $\mathbf{V} = (\mathbf{K}^T \mathbf{K} + \alpha \mathbf{L}^T \mathbf{L})^{-1}$. Where $\mathbf{L}, \alpha$ are the regularization matrix and strength respectively. The regulation strength can be calculated [19] as: $\alpha = \delta / C$, $\|\mathbf{Lx}\| \leq C$ $where \|.\|$ is the norm. Eqn(6) can be modified for highly ill-conditioned Jacobians as

$$H_{hs} = \frac{1}{2} \log_2 \frac{\left| \mathbf{K_2}^T \mathbf{K_2} + \dfrac{\delta}{C} \mathbf{L}^T \mathbf{L} \right|}{\left| \mathbf{K_1}^T \mathbf{K_1} + \dfrac{\delta}{C} \mathbf{L}^T \mathbf{L} \right|} \qquad (7)$$

A precise technique for calculating of the regularization strength is required for a severely ill-conditioned Jacobian. The region of the hyperspace is also dependent on the nonlinearity of the problem.

## Information Analysis

To calculate the information, a model is required. A representative Mars atmosphere is taken from [20]. 95.32% abundance of $CO_2$ has been assumed. The HITRAN-2004 [21] spectroscopic database of the gases is used and the extinction calculation for the gases has been made by the GENSPECT program [22]. We have considered the resolution of the Fourier Transform Spectrometer to be 0.02 $cm^{-1}$ and the geometry of the measurement is solar occultation. The signal-to-noise ratio (SNR) is 100, which is somewhat lower than could reasonably be achieved in practice. The effect of refraction, which is very small because the Martian atmosphere is less dense than Earth, is not considered. We consider a prototype model from 4-50km, which contains 16 grid points. The tangent points are [4, 6, 8, 10, 12, 14, 16, 18, 20, 22.5, 25.0, 27.5, 30.0, 32.5, 35.0 and 40.0] km.

The search space of the proposed problem is wide; for example the atmospheric pressure varies from 0.1pa to 600pa and the temperature varies from $60^oK$ to $300^oK$. If we consider a logarithmic scale of the state space parameters, then the range of the search space is from -2.3 to 6.2 for pressure and from 4.1 to 5.7 for temperature. For this reason we have made this calculation considering the state space as logarithmic.

The length of the tangent level sub-path is very large as compared to other sub-paths in a solar occultation measurement, which restricts the opportunity to extract any information from other sub-paths using any retrieval scheme. For the reduction of computational cost, we consider one signal value for one solar occultation spectrum by integrating the transmittance of a micro-window as $\sum_{i=1}^{N_c}(1-\tau_i)$ where $N_c$ is the numbers of channels in the micro-window and $\tau$ is the transmittance. This will also reduce the nonlinearity and increases the SNR ($800 \approx 10^3$) component without altering the spectroscopic properties in the retrievals.

We select 140 MWs with a width of 1.3 $cm^{-1}$ (64 channels of the resolution of 0.02$cm^{-1}$) from different parts of the spectrum to understand the problem. These windows are basically chosen intuitively from the weak line areas of $CO_2$ bands to produce reasonable signals for the present solar occultation measurements and are presented for illustration purposes only. The path length of the solar occultation geometry is very high (~100km) and measurements at low tangent heights will saturate if the line-strength of the gas lines are high. First, we calculate the two separate Jacobians of pressure and temperature using two different pressure and temperature profiles. One of these profiles is the standard pressure and temperature profile of Mars referred to above. The second profile is a constant profile: T=280K, p=80Pa.

First we have calculated the information content ($H_b$) using most popular method, which is based on the Bayes theory [5] :

$$H_b = -\frac{1}{2}\ln|\mathbf{I} - \mathbf{A}| \qquad (8)$$

where $\mathbf{A} = (\mathbf{K}^T \mathbf{S}_e^{-1} \mathbf{K} + \mathbf{S}_a^{-1})^{-1} \mathbf{K}^T \mathbf{S}_e^{-1} \mathbf{K}$.

Where $\mathbf{S}_a^{-1}$ and $\mathbf{S}_e^{-1}$ are the a-priori and error covariance matrices respectively. The inverse method is so called the optimal estimation method (OEM).

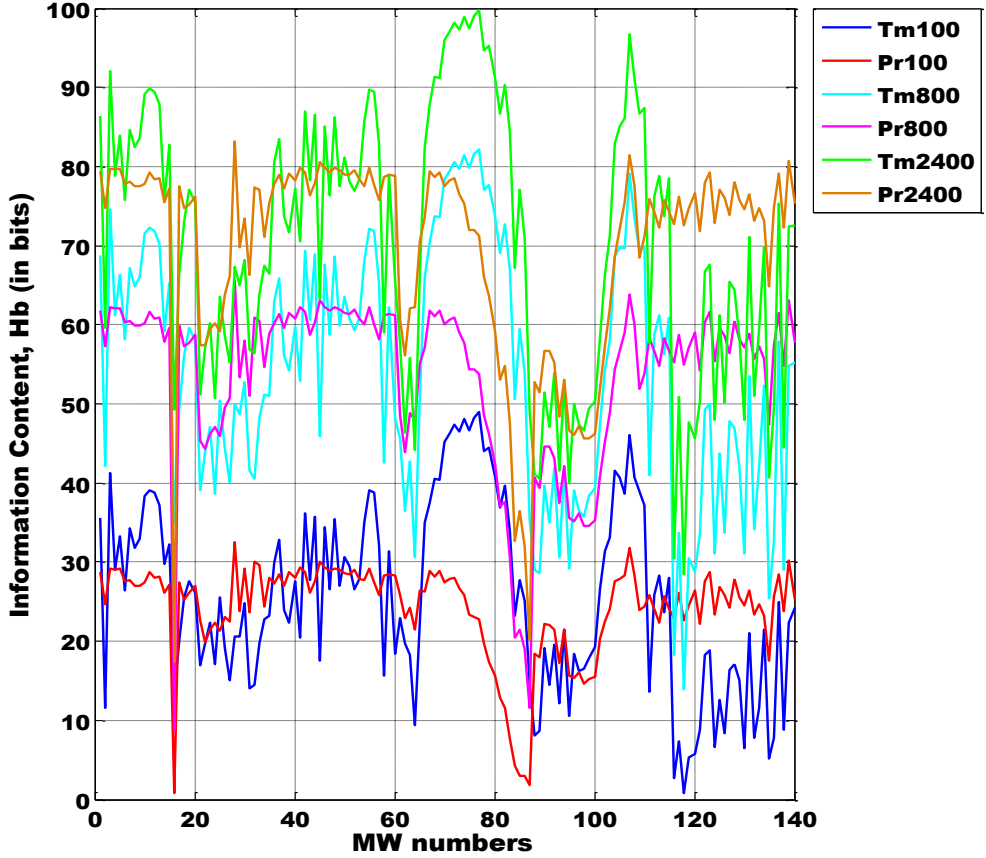

Fig. 1 Information content using Bayes theory for 140 different micro-windows: Tm100 stands for temperature information at SNR=100, similarly Pr800 stands for pressure information at SNR=800.

Since we do not know the proper a-priori covariance, a 50% diagonal covariance is used for these calculations and three different SNRs (100,800,2400) to understand the problem. Fig. 1 shows the values of $H_b$ to be in the range of 20~80 bits (SNR=800) and 1~50 (SNR=100). The values are unreasonably large. According to Shannon [18] in the parameter space, one parameter can produce one bit of information and N parameters N bits of information since the total number of possible states is $2^N$ and $\log_2(2^N) = N$. It should be mentioned that this is valid when the number of measurements greater than or equal to the state space parameters. In our example N=16, thus we expect a maximum of 16 bits of information whereas the results in Fig. 1 show values up to 100. The information content using the same theory of 60 bits has been reported by Dudhia et al. [23], where the problem is almost similar to us (e.g. limb view spectroscopic measurement with a set of parameters of 16) except SNR of this problem is low.

To understand the problem, we explore the basic formulation of this method [5], where the posterior covariance is based on the expected total error $\langle \xi \rangle$

$$\langle \xi \rangle = (\mathbf{y} - \mathbf{Kx})^T \mathbf{S}_e^{-1} (\mathbf{y} - \mathbf{Kx}) + (\mathbf{x} - \mathbf{x}_a)^T S_a^{-1} (\mathbf{x} - \mathbf{x}_a) \qquad (9)$$

Where $\mathbf{S}_e^{-1} and \mathbf{S}_a^{-1}$ are the error and a-priori covariance's, which are effectively scaling factor for the calculation. When the system is perfectly linear and we know the truth exactly, it is possible to approximate error as $((\mathbf{y} - \mathbf{Kx})^T \mathbf{S}_e^{-1} (\mathbf{y} - \mathbf{Kx}))$. The problem is that these problems are not linear. In that case the measurement errors cannot be constructed by using $(\mathbf{y} - \mathbf{Kx})$ because the residual will contain terms from the nonlinear nature of the problem and will therefore not properly represent the measurement errors. Thus, an information calculation using this formulation for any nonlinear problem will produce wrong results. The error injection due to the nonlinearity will be proportional to the nonlinear contribution term multiplied by SNR. At this point we can mention the work of Ceccherini et al [31] where it is stated that $\chi$-square does not exist for nonlinear functions.

Spectroscopic measurements generally deal with SNR of $10^2$-$10^3$ and for the present problem $\sim 10^3$ (corresponding to the single point measurement $\sim 10^2$). The nonlinear contribution of the present problem from the term $(\mathbf{y} - \mathbf{Kx})$ is the order of 0.2. The amount of error inserted in such calculation is the $\log_e(.2 \times SNR) = \log_e(200) \approx 5$. We observed in Fig. 1 that the calculated value of $H_b$ ($\sim 80$ at SNR=800) is approximately 5 times larger than the number of parameters (16). Thus the measurement error statistics or the entropy in measurement from this calculation are corrupted by the nonlinear contribution. This is the source of the problem. By the same argument, even in the linear case, if the residual due to the a-priori assumption is more than the measurement noise, it will produce erroneous results. This is not seen only our problem, the same unreasonably high information content has been reported in many papers [24-30], which confirms that the radiative transfer problem is nonlinear.

The question will arise at this point that there are some successful retrieval have been produced using OEM method, where theory is not appropriate for radiative transfer inverse problem. This can be discussed that the successful retrieval is possible when the choices of $\mathbf{S}_e^{-1}$ and $\mathbf{S}_a^{-1}$ are optimally regularized the problem. To validate our statement, we analyse OEM formulation in the framework of a regularized least squares paradigm. Under such condition, the gain matrix $(\mathbf{G}_y)$ in the Optimal Estimation Method (OEM) can be treated as the inverse of the regularized Jacobian. $\mathbf{G}_y$ is then given by

$$\mathbf{G}_y = (\mathbf{K}^T \mathbf{S}_e^{-1} \mathbf{K} + \mathbf{S}_a^{-1})^{-1} \mathbf{K}^T \mathbf{S}_e^{-1} \qquad (10)$$

Usually the error covariance is diagonal and the individual terms the same. In such cases, $S_e$ is a scalar term and $G_y$ can be rewritten as: $G_y = (K^T K + S_e S_a^{-1})^{-1} K^T$. Under the regularized least squares assumption[32,33] the information matrix is $(K^T K + S_e S_a^{-1})$ and the posterior covariance is $S_p = (K^T K + S_e S_a^{-1})^{-1}$. Using this formulation the information can be calculated as

$$H_{bm} = -\frac{1}{2} \ln \left| S_p S_a^{-1} \right| \qquad (11)$$

Where $H_{bm}$ is the modified information content using Bayes theory. The calculated $H_{bm}$ for our 140 MWs for the standard pressure and temperature profiles of Mars with three different 100%, 50% and 10% diagonal a-priori covariances for SNR=800 are presented in Fig. 2. Notice that the calculated information has dropped significantly, but is still high.

The modified calculation of information ($H_{bm}$) shown in Fig. 2 demonstrates that the values of $H_{bm}$ can be within the domain of the number of parameters (16) for a small a-priori covariance of 10%, whereas even if we recomputed Figure 1 with 10% a-priori covariance instead of 50%, the values of $H_b$ would still be in the range 50~140. This also confirms that the regularization paradigm can produce better results than the error statistics estimation approach for a nonlinear problem. The values of $H_{bm}$ very much depend upon the choice of the a-priori covariance and the calculated "information" increases with increasing a-priori covariance. This implies that a-priori covariance is obscuring the information in the measurement. It is very difficult to partition the information between the measurement and the a-priori in such a formulation.

Fig. 2 shows that the calculated maximum information content of temperature-sensitive MWs can be up to double the maximum information content of pressure-sensitive MWs. This can be understood in terms of the functional nonlinearity. This problem is not linear and has a complex functional behaviour. It is not easy, or perhaps not even possible, to determine the quantitative degree of nonlinearity. On the other hand, when the covariance calculation is based on the model, the functional relation of the model is important and we have assumed a linear relation of the forward model to develop the covariance in the information content calculation in Eqn(6). If the assumption is very far from reality, the information content calculation will almost certainly lead to erroneous results.

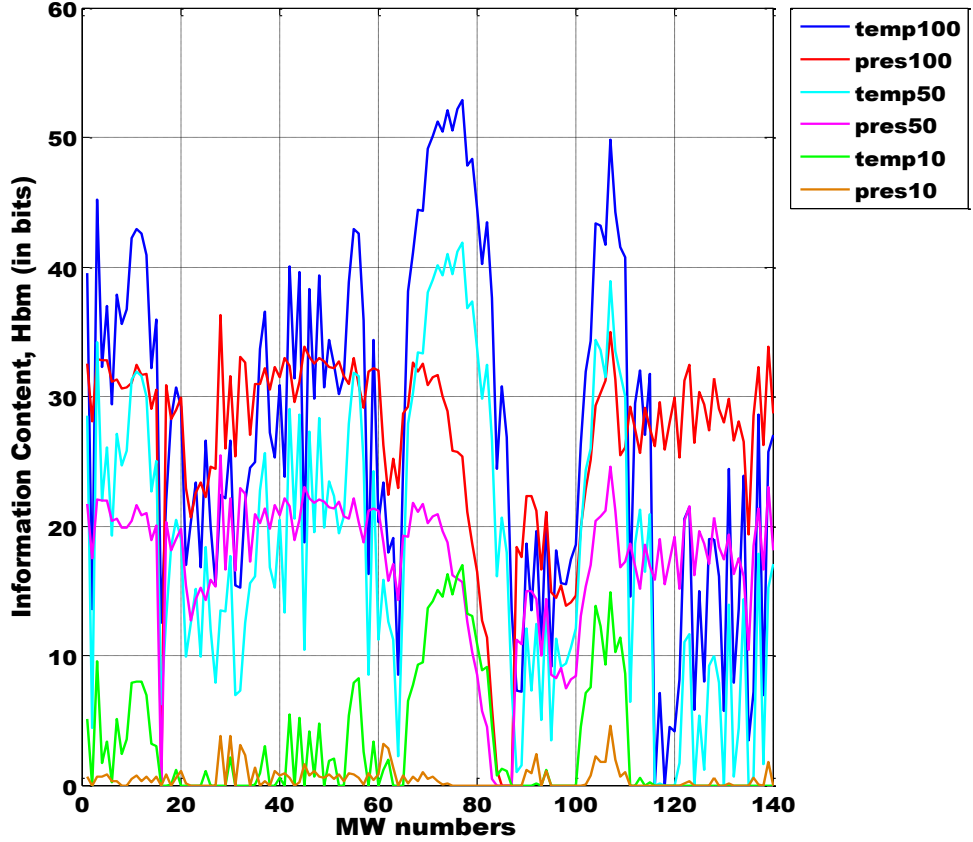

Fig. 2 The information of $H_{bm}$ for pressure and temperature under three different a-priori covariance (i.e. temp100-> 100% a-priori covariance of temperature; similarly pres50->50% a-priori covariance of pressure).

We have made a simple trial to depict the effect the degree of nonlinearity of the present problem. We first select for illustration purposes two MWs from Fig. 2. For the first the temperature information is very high (MW-79) and for the second one the temperature information is low (MW-132). We then calculate the signal using the standard Mars atmospheric pressure ($\mathbf{P}$) and temperature ($\mathbf{T}$), which is plotted in Fig. 3. This implies $\mathbf{x} = [\mathbf{T}; \mathbf{P}]$. We have calculated the Jacobians of these MWs for temperature and pressure $\mathbf{K}_{p1}$, $\mathbf{K}_{t1}$, $\mathbf{K}_{p2}$ and $\mathbf{K}_{t2}$ Then, we plot four different functional relations as follow:

$[\mathbf{K}_{t1}\ \mathbf{K}_{p1}]\mathbf{x}$, $[\sqrt{\mathbf{K}_{t1}}\ \mathbf{K}_{p1}]\mathbf{x}$, $[\mathbf{K}_{t1}\ \sqrt{\mathbf{K}_{p1}}]\mathbf{x}$ $and$ $[\sqrt{\mathbf{K}_{t1}}\ \sqrt{\mathbf{K}_{p1}}]\mathbf{x}$ in Fig. 3a .

$[\mathbf{K}_{t2}\ \mathbf{K}_{p2}]\mathbf{x}$, $[\sqrt{\mathbf{K}_{t2}}\ \mathbf{K}_{p2}]\mathbf{x}$, $[\mathbf{K}_{t2}\ \sqrt{\mathbf{K}_{p2}}]\mathbf{x}$ $and$ $[\sqrt{\mathbf{K}_{t2}}\ \sqrt{\mathbf{K}_{p2}}]\mathbf{x}$ in Fig. 3b.

That is, we have compared a linear relationship and a square root (non-linear) relationship.

(The matrix notation is

$$[\mathbf{A}\,\mathbf{B}] = \begin{bmatrix} a_{11} & a_{12} & ... & b_{11} & b_{12} & ... \\ a_{21} & a_{22} & ... & b_{21} & b_{22} & ... \\ ... & ... & ... & ... & ... & ... \end{bmatrix}$$

Where $\mathbf{A}\,\&\,\mathbf{B}$ are two different matrices with elements $a_{11}, a_{12}, ...$ & $b_{11}, b_{12}, ....$ respectively.)

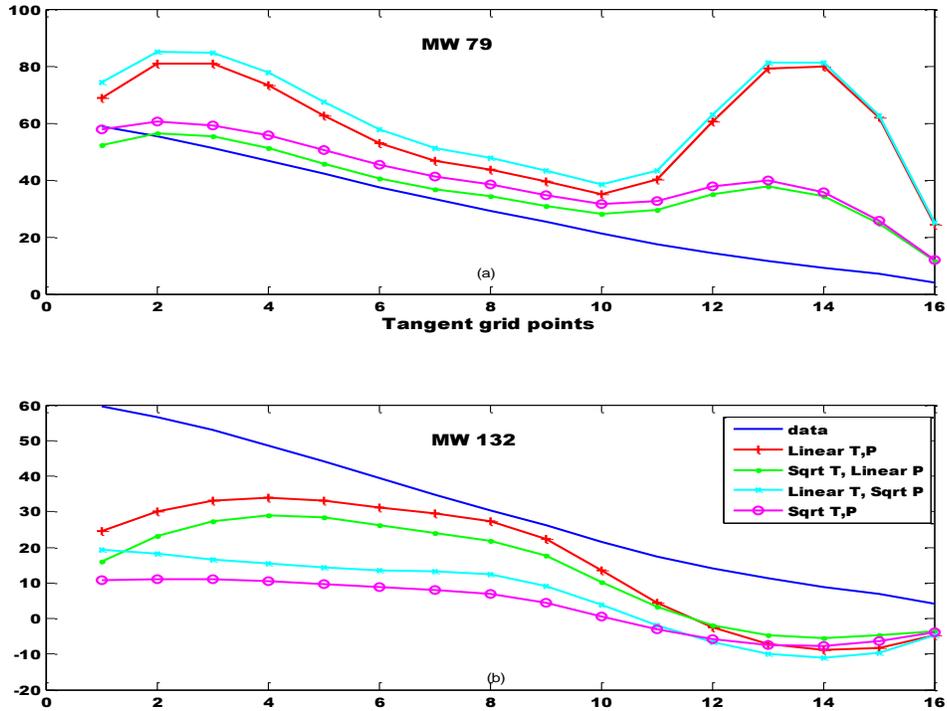

Fig. 3 Verification of the degree of nonlinearity of present model using a linear and a square roots of Jacobians.

If the functional relationship is correct, then the appropriate line in Fig. 3 should lie on top of the data, i.e. in that case [... ...]x=y. The further the functional line departs from the data, this means that the assumed functional relationship departs from the actual functional relationship. We categorised these in four groups as: good, reasonably good, poor, very poor.

It is very difficult to develop a simple nonlinear relation of the present problem. The nonlinearity pattern is very complex and highly dependent upon the MWs selected since the nonlinearity patterns are very different for various parts of the spectrums and different tangent points. Table-1 shows that the square root of temperature jacobian and a linear pressure jacobian produces the best overall relationship and can at least be used as a qualitative measure.

Table - 1 Proximity of the functional relationship

|  | MW-79 | MW-132 |
|---|---|---|
| $[K_t \ K_p]$ | poor | good |
| $[\sqrt{K_t} \ K_p]$ | good | reasonably good |
| $[K_t \ \sqrt{K_p}]$ | very poor | poor |
| $[\sqrt{K_t} \ \sqrt{K_p}]$ | reasonably good | Very poor |

We observed the condition number of these jacobian $O(10^2)$, which is not highly ill-conditioned and we can use the Eqn(6) for calculating the information using hyperspace. Since the present problem produces a triangular matrix, $\left|\sqrt{\mathbf{K}}\right| = \sqrt{|\mathbf{K}|}$. we can modify Eqn(6) as

$$H^t_{hs} = \frac{1}{4} \log_2 \frac{\left|\mathbf{K}^T_{2t}\mathbf{K}_{2t}\right|}{\left|\mathbf{K}^T_{1t}\mathbf{K}_{1t}\right|} \quad and \quad H^p_{hs} = \frac{1}{2} \log_2 \frac{\left|\mathbf{K}^T_{2p}\mathbf{K}_{2p}\right|}{\left|\mathbf{K}^T_{1p}\mathbf{K}_{1p}\right|} \qquad (12)$$

Where, $H^t_{hs} \ and \ H^p_{hs}$ are the information content of temperature and pressure using hyperspace respectively. The notation of the Jacobian: $\mathbf{K}_{1t} \ and \ \mathbf{K}_{t1}$ stands for the temperature Jacobian for different profiles and different MWs respectively.

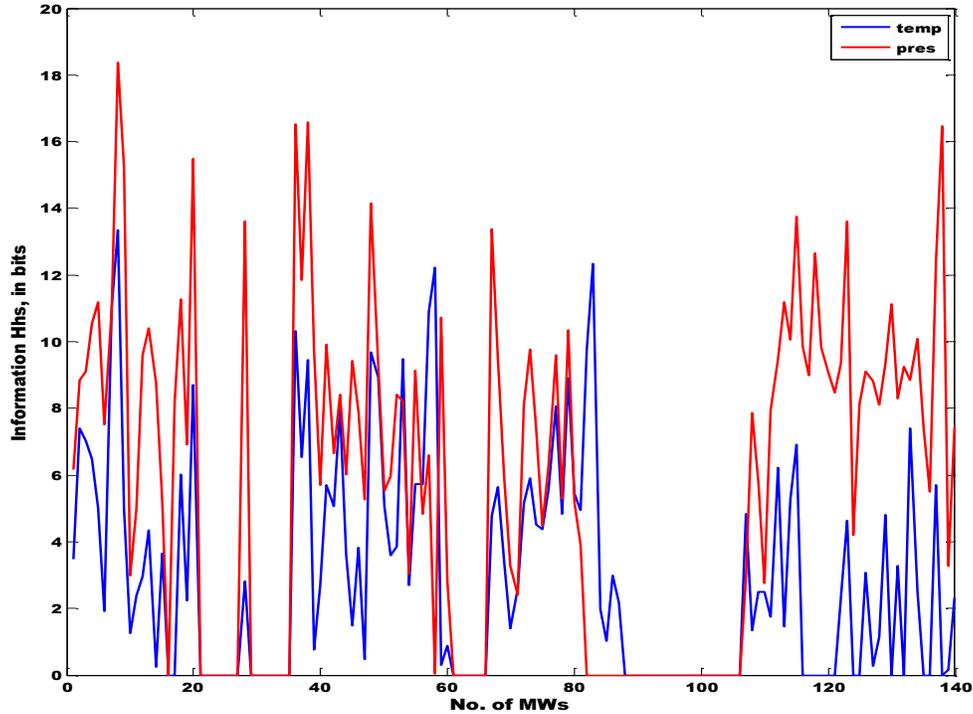

Fig. 4 Information content using hyperspace of 140 different Micro-windows.

Using all of the above, we compute Figure 4 realising that it is only a rough first-order approximation to a very complex problem. However, even with this crude calculation, the numbers do finally become reasonable. The number of bits of information is of the same order as the number of parameters. The calculated information for the pressure component in Fig. 2 for 5% a-priori is quite similar in nature as we obtained in Fig. 4 using hyperspace formulation, because in that case the relationship is linear.

## *Micro-Window (MW) Selection*

There is no doubt that the pressure and temperature retrieval from this spectroscopic measurement is an ill-posed problem. Therefore it is impossible to solve the problem at an arbitrary location in the spectrum. However, there could be some micro-windows (MWs) available where the problem can be solved. Thus, we need to select appropriate MWs by considering both the physics and mathematics of the problem.

Consider the following: the signal is primarily a function of the number density of molecules of the species being considered at a level. The same density may be found for several sets of pressure and temperature for a constant volume mixing ratio since they are linked via the ideal gas law. In such a situation, we have to be very careful to select MWs in such a way that there is maximal information content for a particular set of parameters (e.g. the pressure/temperature profile points) together with a minimal information for another set of parameters (the temperature/pressure profile points). Alternately stated, the mutual information between two sets of parameters has to be minimized. This implies that there should be a large difference between the information for the two sets of parameters. Ideally the value of the information of one set of parameters should be zero and the value of the other very high.

For example, if $H_{hs}$ for temperature is 10 and for pressure <1 for a specific MW, then we can choose that for the temperature retrieval. With such parameters, the MW will be temperature sensitive but pressure insensitive. We have to select another MW using the same concept for the pressure retrieval. It can be seen from Fig. 4 that a limited number of pairs of MWs can produce such a combination. However it is this combination that might lead to a unique solution for the pressure and temperature retrieval.

As stated above, the main criteria of the MW selection is to minimize of the mutual information among the two sets of variables, but there are some other problems associated in this selection process such as the gradient of consecutive measurements and the error from the interfering gases. When the gradient is flat, as when the measurement is saturated, it is difficult to find a reasonable solution. To avoid errors from other interfering gases, we can use the select-channel model, which is discussed in ref.[7]. Following the above mentioned criteria, we have selected the MW 118 for the pressure retrieval where the error due to interfering gases < 0.2% and MW 83 for the temperature retrieval. We use these MWs in the following case study. The information in bits of temperature and pressure are T~0, P~8 (MW-118) and T~8,P~0 (MW-83).

## *Synthetic Data Generation*

The most significant difference between the forward model for Mars and that for Earth is the dust. Dust scattering is the most important extinction process in the Martian atmosphere, and must be considered in synthetic data preparation. The extinction coefficient at optical wavelength λ due to dust, $\sigma_\lambda$ (m$^{-1}$) can be modelled as

$$\sigma_\lambda = \int_0^\infty N(r,z)\, k_{r,\lambda}\, dr \qquad (13)$$

where $k_{r,\lambda}$ is the effective scattering cross-section of a single dust particle of radius r. For spherical dust particles, the effective scattering cross-section for a single particle is given by

$$k_{r,\lambda} = \pi r^2 Q_{r,\lambda} \qquad (14)$$

where $\pi r^2$ is the actual dust particle cross-sectional area, and $Q_{r,\lambda}$ (dimensionless) is the single particle scattering efficiency. The scattering efficiency is a function of a complex index of refraction, the wavelength of the light and a size parameter. We have collected the extinction efficiency data for 0.3~4.15 μm from Ockert-Bell et al. [34] and 6~16 μm from Forget [35] . No dependence on particle size is mentioned, but the effective diameter of particle is considered to be 1.85 μm. The optical depth (dimensionless) quantifies the scattering and absorption that occurs between the top of the atmosphere can be calculated as

$$\tau_\lambda = \int \sigma_z dz \qquad (15)$$

For $Q_{r\lambda} = Q_{ext}$, independent of r, and thus of height, z, we can write

$$\tau_\lambda = Q_{ext} \int A(z) dz \qquad (16)$$

where A(z) is the cross-sectional area per unit volume

$$A(z) = \int_0^\infty \pi r^2 N(r,z) dr \qquad (17)$$

We need a dust profile to do our numerical simulation. We used a simplified dust profile with a scale height (H) of 10km and uniform distribution. Then we can write

$$\tau_\lambda = Q_{ext} N_0 \int \pi r^2 e^{-\frac{z}{H}} dz \qquad (18)$$

Where $N_0$ is the dust concentration at the ground. The value of $N_0$ of 1.85 μm particles at 1075cm$^{-1}$ for a vertical optical depth of 0.5 is 1.68x10$^6$ m$^{-3}$. If we assume particle size of 1 μm then it calculates $N_0$ of 5.68x10$^6$ m$^{-3}$ and the column particle integrated cross section 5.9x10$^{10}$ m$^{-2}$. These values appear reasonable as the Pheonix dust model [36] calculates a number density of 1.47x10$^7$ m$^{-3}$ and  the column particle integrated cross section of $5.4 \times 10^{10}$ m$^{-2}$. The dust on Mars is known to be highly variable, so these values should be treated as approximations.

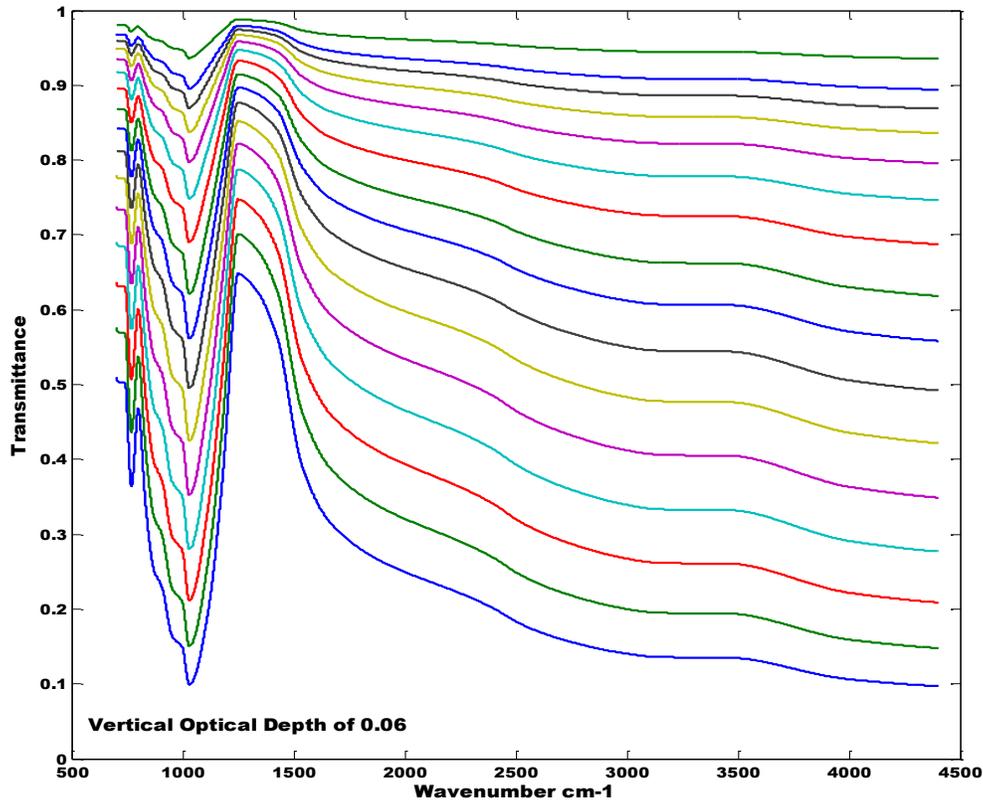

Fig. 5 A calculated dust transmission spectra for various tangent height of a solar occultation measurement in low dust conditions.

Figs. 5 & 6 show the transmission of dust for two different vertical optical depths 0.06 and 0.5. It is observed that the proposed solar occultation instrument cannot make any measurement below 25km when the vertical optical depth of dust is as high as 0.5.

We will now present a prototype retrieval from 4-50km, which contains 16 grid points, with a dust vertical optical depth of 0.06. To simulate the data, we added Gaussian random noise of mean zero and standard deviation of 0.01 produce a SNR=100 for 100% transmission. Simulated retrievals have been performed using a regularised total least squares method (RTLS), which is described in ref [7-8]. Three runs of each retrieval have been done using three different realizations of the noise to gain confidence in the retrieval scheme. We solve the problem by considering both sets of parameters as a single vector (e.g. [T;P]$^T$ means [t$_1$ t$_2$ ...; p$_1$ p$_2$...]$^T$). We constructed a synthetic "true" profile in such a way that there is a sinusoidal variation of amplitude of 25% around the standard temperature and 50% around the standard pressure profile of the Mars. We have performed our simulated retrieval from two different initial guess profiles: the standard pressure and temperature profile of Mars referred to above and a constant profiles of T=280K, p=80Pa. The present simulated retrieval assumes that the differences between the tangent heights of the individual spectra are known, but not the absolute value. This corresponds to the case of a spacecraft where the relationship between the spectra taken in a single occultation is well-known from spacecraft and scanning parameters, but a fiducial value related to either the atmosphere or the planet has to be obtained from the data.

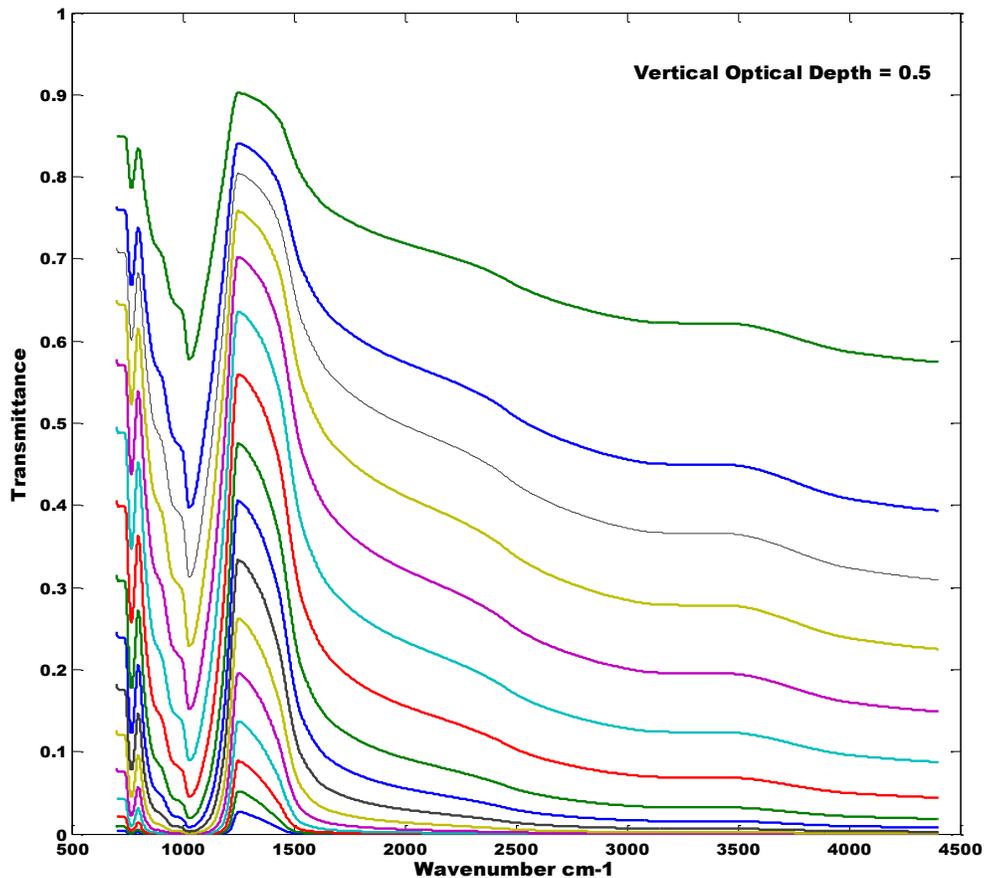

Fig. 6 A calculated dust transmission spectra for various tangent heights of solar occultation measurement in moderate dust conditions.

## *Results and Discussions*

The common belief is that the information is the measure for the success of an ill-posed inversion. According to Fig. 1, the information ($H_b$) of both parameters is quite high (~160 bits) for MWs 1~15 and/or 40~50 as compared to the other MWs. Thus, one can expect that there will be a reasonable solution for any combination of these MWs. However, we did not get a reasonable solution using OEM showing that the present problem cannot be solved using any MWs rich in information for both sets of parameters, even if these MWs are supplemented with a-priori information and forced by a-priori constraints. The a-priori information and/or constraint are basically noise in terms of the true solution process as is discussed also in ref [6,8]. The solution process is governed by the information in the measurement, the condition number of Jacobian, the functional complexity, prevention principle of the multiple solutions and the stabilizing criteria (that blocks the noise propagation) .

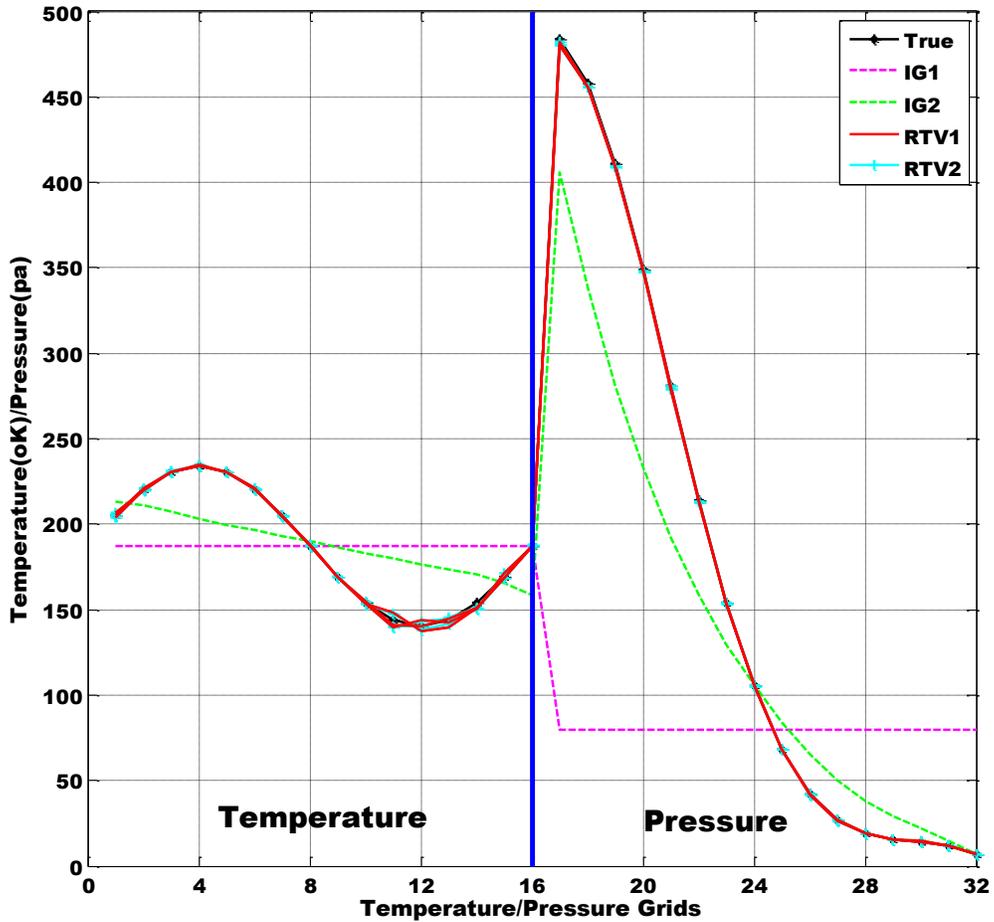

Fig. 7 A composite retrieval of pressure and temperature from a spectroscopic measurement. The grids 17 to 32 correspondence the pressure grids 1 to 16.

The retrieval results are shown in Fig. 7, showing that there exists a unique solution of these parameters from the spectroscopic measurement using this methodology. There is a little oscillation observed when both temperatures and pressure are low, however, the error is less than 2% . The signal is very low at the low end of the temperature and pressure grid and noise is corrupting the signal. This problem can be minimized by selection of several appropriate MWs for several ranges of P,T. It is often argued that the regularization suppresses the information to smooth out the solution. To demonstrate otherwise we have plotted the pressure and temperature as a single vector (as it is solved) to emphasize that our regularization scheme does not do this. The solutions tend to the correct result even with the big difference between the last temperature point and of the first pressure point. This is possible because the RTLS method implicitly decreases the regularization strength when the iterative process approaches the final solution. Thus the regularization effect close to the solution point is very low and all the information is retained in the solution.

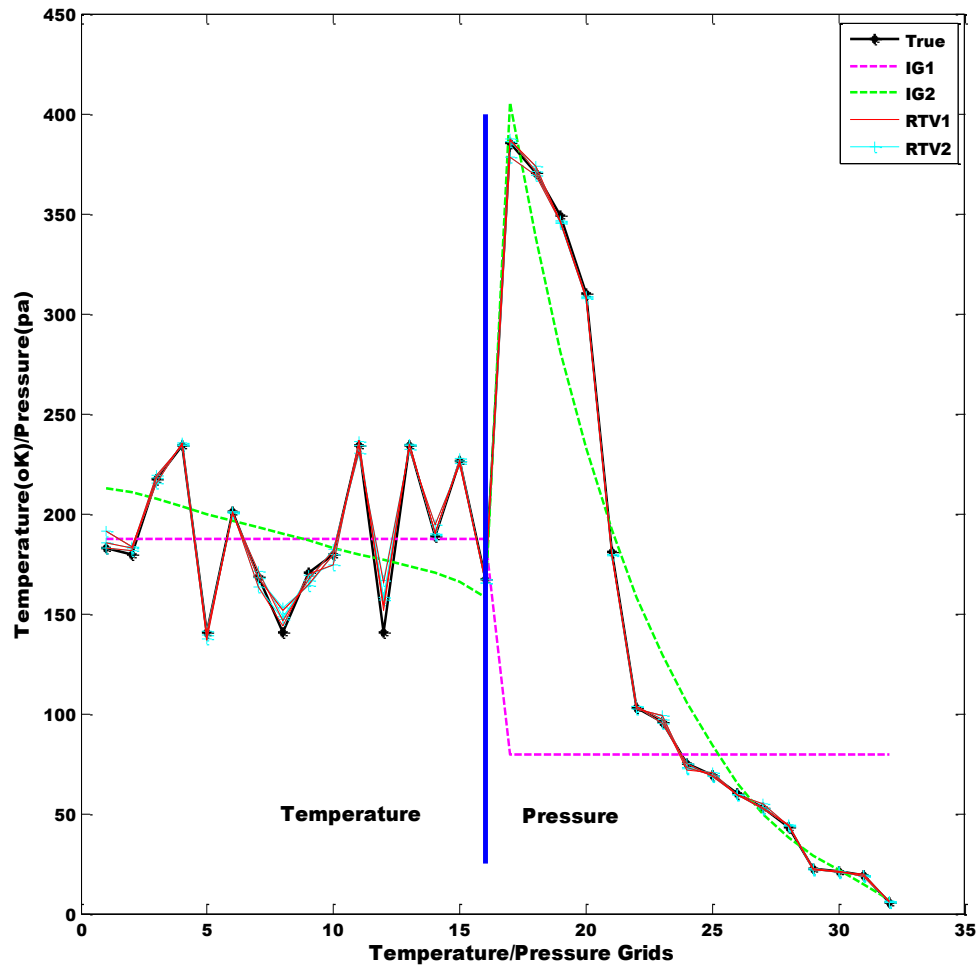

Fig. 8 A composite retrieval of pressure and temperature for random profiles. The grids 17 to 32 correspondence the pressure grids 1 to 16.

To further verify the regularization effect in our retrieval schemes, we have assumed the truth is not a smooth profile. The true profile is the standard profile modified at random by up to 25% in temperature and 50% in pressure. The results are shown in Fig. 8. The results of this study gives an assurance that the pressure and temperature can be uniquely solved using a spectroscopic measurement. In this case the low temperature points at high pressures do not produce any oscillation.

A issue can be raised that the information calculation using the hyperspace formulation is independent of the noise in the measurement. In considering this point it can be argued that the noise in the measurement is not the limiting factor (i.e. our experiment can be achieved at SNR~500), the major issue is the ill-conditioning of the problem. We have examined this issue by calculating the information using Eqn(7) and Eqn(11). The difference is very small due to the inclusion of the error terms in the calculation. We have already discussed that the information calculation using the entropy method is a qualitative measure whereas the problem is solved by an iterative method. The value of degree of freedom of retrieval (DFR), which is discussed in ref[6], at the last iteration contains the meaningful information for the inversion. The basic idea behind the information analysis using hyperspace is the understanding of the physical problem in perfect measurement conditions in order to select the optimum MWs.

Yet another issue is why we use a high resolution measurement when 64 spectral points are co-added to produce the "signal". The high resolution measurement is necessary so that we can eliminate the unwanted features from the spectrum using our select channel model [7] before adding the spectral points. The advantages of adding spectral points are a reduction in the problem nonlinearity as well as the measurement noise without a corresponding disturbance of the spectral features and information.

There is no dependence upon the initial guess: Our retrieval scheme produces the same solution for two very different initial guess profiles as shown in Figs. 7 & 8. Our solution does not produce any error correlated with the error in the fiducial tangent height. An error of 5km in tangent height will introduce only $7.3*10^{-2}$ % error in our solution. This happens because we do not use a hydrostatic constraint in our retrieval model. We do require that we know the relation between successive measurements, which is easier to determine from a satellite based measurement as discussed in the introduction.

Despite the fact that the proposed retrieval scheme gives a good result, there still remains the problem of the fidelity of our dust model. Our dust model is based on the single scattering and only silicate as a scattering material, which is a reasonable assumption for the Martian dust. We now present a variant of our retrieval scheme that avoids the ambiguity of dust. We assume that the transmittance of the dust is fairly constant within the small portion of spectrum in a MW and scale the calculated transmittance (with noise) with respect to the maximum value in the MW before adding the spectral points. The advantage is that the scaled retrieval will remove any other broad band features. As we assume that the detector is dominated by "Johnson noise" in our simulated retrieval (worst case scenario − photon noise limits would produce a better result), the noise is enhanced 4~5 times for lower tangent height measurements due to the high absorption of the dust. It should be mentioned that the assumed SNR=100 of this study, which is somewhat lower than could reasonably be achieved in practice. SNR=400 could probably be achieved with this measurement technique[12]. Thus, we have made same retrieval using SNR=300 without the dust model in the retrieval scheme, where as the synthetic data contains the dust transmittance. The optimized MW will change when the model is changed. In such a situation, we intuitively select the optimized MWs for this study, the details are not shown here.

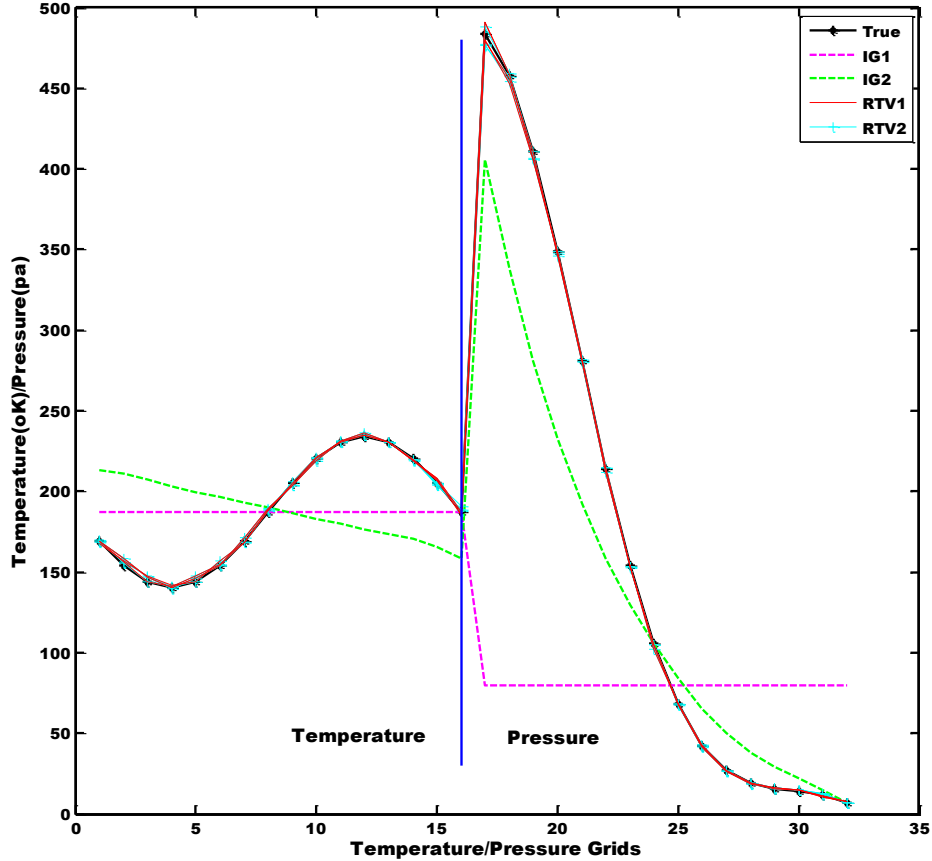

Fig. 9 A composite retrieval of pressure and temperature without dust model in retrievals. The grids 17 to 32 correspondence the pressure grids 1 to 16.

The results as shown in Fig. 9 confirm that the ambiguity of dust in Martian atmosphere can be avoided for the pressure and temperature retrieval by using our model. The present retrieval scheme has stable solutions even under scaling. The solution contains more error due to the high noise in the input signal (the simulation under SNR=100, which is not shown here).

We have also made a similar simulated study for the satellite based solar occultation measurement of the Earth atmosphere (results are not shown here). It is easier to solve the problem as the dust is not present in Earth atmosphere. We find that the pressure and temperature of Earth's atmosphere can be uniquely retrieved from such spectroscopic measurements using this method.

As a final test to authenticate the present retrieval scheme we now construct 25 random profiles of temperature and pressure around the Martian standard temperature and pressure profile as shown in Fig. 10.

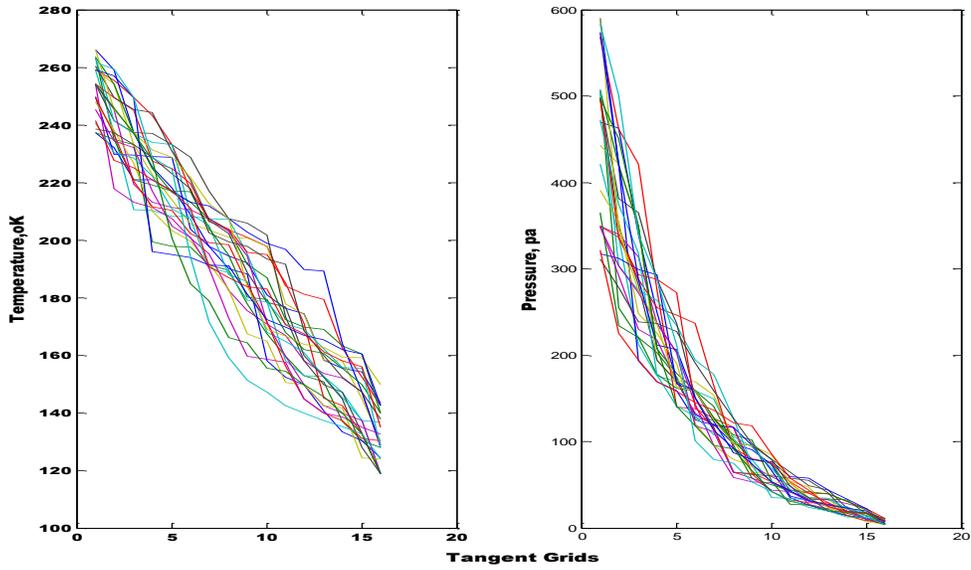

Fig. 10 Twenty five random profiles of temperature and pressure around the Martian standard profile.

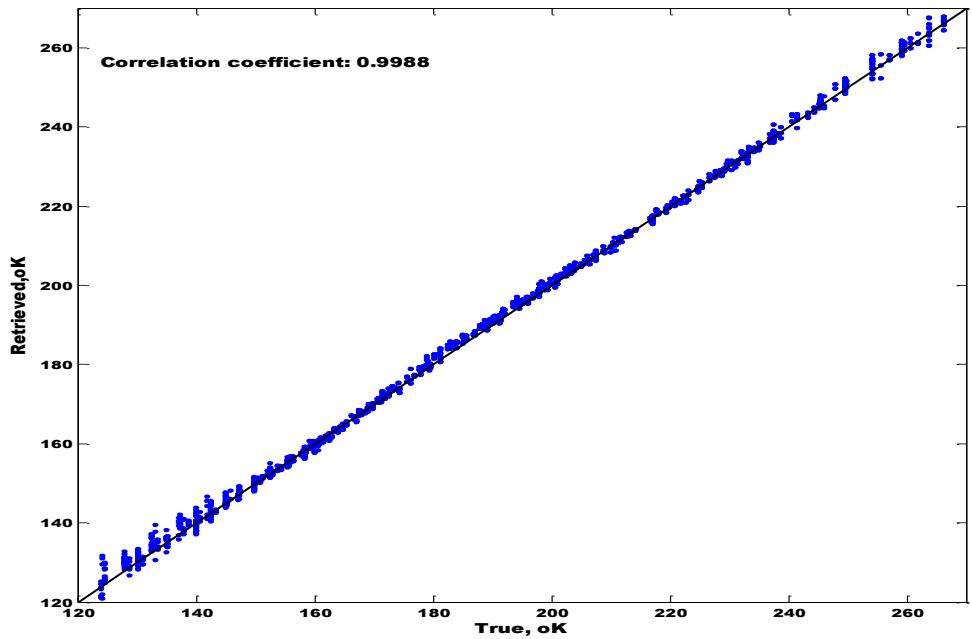

Fig. 11 Scatter plot of temperature retrieval of 25 random profiles.

The problem is then solved by the second option: the dust transmittance is included in synthetic data but the retrieval is made using scaling, without the dust model and SNR=300. We used two different initial guess profiles and three realizations of random noise. The scatter plots of temperature and pressure (Figs 11 & 12) each have 2400 points, enough to do statistical inference.

The correlation coefficient of present retrievals for the temperature and pressure are 0.9988 and 0.9999 respectively. This confirms that the pressure and temperature can be uniquely mapped from the proposed measurement. The retrieval of temperature below 140K shows higher scatter because the signal at low pressure and temperature is very low.

After successful retrieval of the pressure and temperature, and the assumption of 95% constant volume mixing ratio of $CO_2$ in the Mars atmosphere, the optical depth of the dust in the MWs can easily be retrieved as can numerous trace gases.

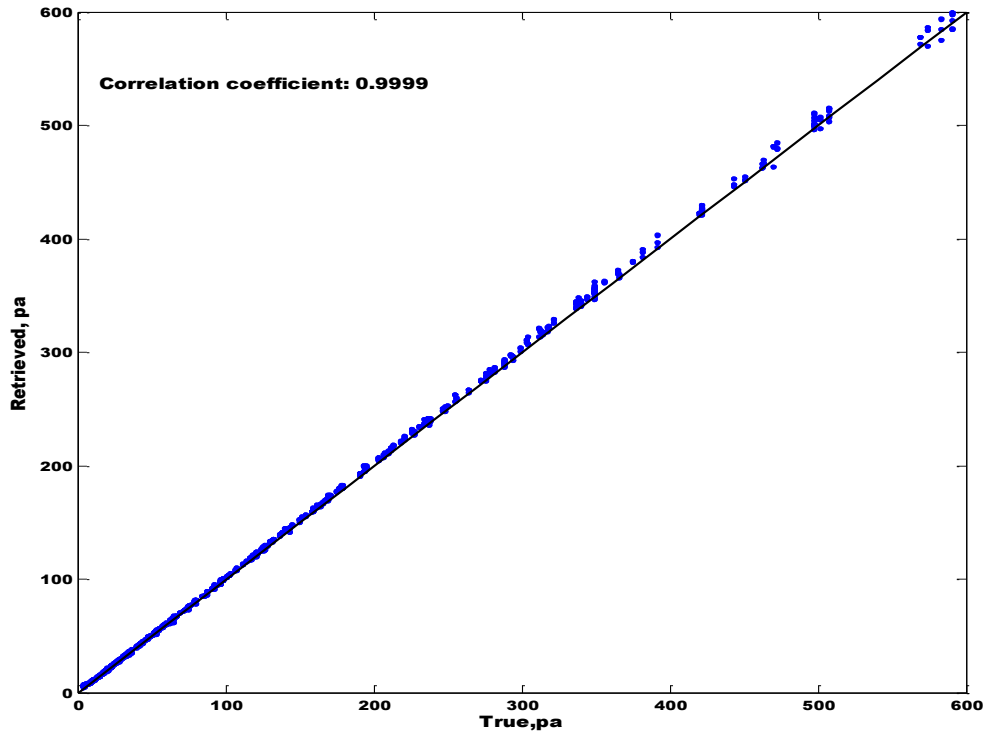

Fig. 12 Scatter plot of Pressure retrieval of 25 random profiles.

## Conclusions

We have demonstrated a method to retrieve the pressure and temperature of the atmosphere of Mars independently and uniquely from a reasonable starting profile in the domain. Our solution is independent of the initial guess profile, a-priori information and tangent height.

A new methodology for the calculation of information in measurement based on the model space parameters has been introduced, which helps to understand the physical problem in an ill-posed inversion. We also include a new technique for the micro window selection for the two sets of parameters (pressure and temperature) using only the information content of the measurements.

The success of this retrieval is due to optimum MWs selection using hyperspace information content analysis. It can be concluded that the information analysis using Bayes formulation has its own inherent limitations: the information in bits are erroneous when problem is nonlinear and a-priori information cannot be partitioned from the measurement. The information rich MWs for both sets of parameters are not necessarily a good choice to achieve a reasonable solution when the physical problem produces multiple solutions. The mutual information between the parameters, where multiple solutions occur, has to be minimized in order to achieve the reasonable retrievals.

We also showed that the effect of the dust in Mars can be successfully eliminated from the pressure and temperature retrievals.

## *Acknowledgement*

We acknowledge the support of the Canadian Space Agency and ABB Bomem Inc. to do this work.